\documentstyle[aps]{revtex}
\input epsf.tex
\def\ba{\begin{eqnarray}}
\def\ea{\end{eqnarray}}

\def\ll{\langle \langle}
\def\rr{\rangle \rangle}

\begin{document}

\title{Quantitative study of scars in the boundary section of the 
stadium billiard}

\author{ Fernando P. Simonotti, Eduardo Vergini, Marcos Saraceno}

\address{Departamento de F\'{\i}sica, Comisi\'on Nacional de Energ\'{\i}a
At\'omica, Av.\ Libertador 8250, 1429 Buenos Aires, Argentina.}

\maketitle

\begin{abstract}
We construct a semiclassically invariant function on the boundary 
of the billiard, taken as the Poincar\'e 
section in Birkhoff coordinates, based on periodic orbit 
information, as an ansatz for the normal derivative of the 
eigenfunction. Defining an 
appropriate scalar product on the section, we can compute the 
{\it scar intensity} of a given periodic orbit on an eigensate, 
as the overlap beetween the constructed function and the normal 
derivative on the section of the eigenstate. In this way, we are 
able to investigate how periodic orbits scar the spectrum and 
how a given eigenstate decompose into {\it scar functions}. 
We use this scheme on the Bunimovich stadium. 


\end{abstract}

\section{Introduction}

Since the observation of imprints of periodic orbits in quantum 
eigenfunctions, {\it scars}, by McDonald and Kaufmann 
\cite{mcdonald}; a vast 
amount of work has been done towards the understanding of this 
phenomenon. The numerical work and theoretical analysis of Heller 
\cite{heller} has been of great importance. Bogomolny 
\cite{bogoscars} pushed the theory of scars further, his 
developments relied on the smearing of the probability density 
over a small energy range. A similar approach, but in phase space 
rather than in coordinate space, was used by Berry \cite{berryscars}.
A theory for individual eigenstates was developed by Agam and 
Fishman \cite{agam}, who constructed a semiclassical Wigner function. 
The integration of this Wigner function in a narrow tube along a 
periodic orbit gave them the scar weigth. Smilansky \cite{smilansky}
used the scattering approach to define a function on the Poincar\'e 
section, which was tested for scars. An important tool in the 
following analysis, the stellar representation, 
was developed by Tualle and Voros \cite{voros}.

In this paper, we construct a semiclassicaly invariant function 
on the Poincar\'e section, built 
on a given periodic orbit, which we call the {\it scar function}, 
which can be extended to the domain via the Green theorem 
(\ref{green2}). 
We define the {\it scar intensity} as the overlap beetween this 
scar function and the corresponding reduction of the eigenfunction 
on the section, with a given measure, so as to mimic the overlap 
in the domain. Using this construction on the stadium billiard, 
we are able, by means of the symbolic dynamics, to identify scars 
of single periodic orbits and of families of them in the 
quantum spectrum. Also, as we do not resort to energy smearing, we 
can decompose an eigenstate in periodic orbit functions (the scar 
functions.) 

This paper is organized as follows. In section \ref{sfb} we 
formulate our approach to the calculation of a scar measure, and 
introduce the necessary objects. In section \ref{bunis} we 
compute scars for the Bunimovich stadium and analize how 
different families of periodic orbits scar the corresponding 
eigenfunctions, particularly the whispering gallery and bouncing 
ball families. Our conclusions and closing remarks can be found 
in the last section.

\section{ The Scar Function on the Boundary}
\label{sfb}

All the information of a given eigenfunction of the billiard 
(with Dirichlet boundary conditions) is 
contained in its normal derivative evaluated on the boundary. By 
means of the Green function we can obtain the wave function 
$\Psi_{\nu}({\bf r})$, with wave number $k_{\nu}$, in the
domain in terms of its normal derivative:
\ba
\label{green}
\Psi_{\nu}({\bf r})= 
\oint ds ~ G_{0}(k_{\nu};{\bf r},{\bf r'}(s)) ~
{ {\partial \Psi_{\nu} } \over {\partial {\bf n}} } ({\bf r'}(s) )
\ea
where $G_{0}(k_{\nu};{\bf r},{\bf r'}(s))=(-i/4) H_{0}^{(1)}
(k_{\nu}|{\bf r} - {\bf r'}(s)|)$ is the free Green function
\cite{berry}. $H_{0}^{(1)}$ is the Hankel function of the first kind.

We thus treat the normal derivative as the fundamental irreducible
object that will be tested for the presence of scars. The function
\ba
\label{normald}
\phi_{\nu}(s) = 
{ {\partial \Psi_{\nu} } \over {\partial {\bf n}} } ({\bf r'}(s) )
\ea
can be thought of as the coordinate representation of an abstract
vector $| \phi_{\nu} \rangle$ in the Hilbert space of periodic
square integrable functions on the boundary. This boundary is also 
the standard 
Poincar\'e section for the classical dynamics and reduces the motion
to a canonical mapping in the Birkhoff coordinates $(q,p)$ 
 \cite{arnold}. The coordinate $q$ is related to the arc length 
coordinate at the boundary where the bounce takes place by 
$q=(s/\mbox{perimeter})_{\mbox{mod}~ 1}$; and 
$p~=~{\bf \vec p} \cdot {\bf \hat t} / |{\bf \vec p}|$
is the fraction of tangential momentum at this point.  

The Fourier transform of (\ref{normald}) would represent it in the 
momentum representation and a coherent state one would display its 
features in the Birkhoff coordinate plane. We follow \cite{voros} 
to choose this route. Other representations are, of course, possible 
and have been used in this context. Smilansky \cite{smilansky}, 
for example, has used angle and angular momentum variables as 
phase space coordinates, which is the natural description when 
employing scattering methods. 
However, in this basis the representation of diffraction effects on 
functions on the boundary is singular and difficult to observe.
The Birkhoff coordinates, besides embodying the natural geometry of 
the billiard, avoid this problem by prescribing definite periodic 
functions as candidates for boundary eigenfunctions.

As shown in \cite{vergini}, for  
eigenfunctions normalised to unity in the domain, in a 
neighbourhood of a given $k$ value, we have the quasi-orthogonality 
relation for the normal derivatives
\ba
\label{qortho}
{1 \over {2 k_{\nu}^{2}}} \oint ~ ds ~{\bf \hat n} \cdot {\bf r}~ 
\phi^{*}_{\nu}(s) \phi_{\mu}(s) ~=~
\delta_{\mu\nu}~+~
{ {(k_{\mu}-k_{\nu})} \over {(k_{\mu}+k_{\nu})}  } O(1).
\ea
Thus, with this measure, the set of eigenfunction in a narrow range 
of $k$ is orthonormal and span a linear space of dimension 
$O(k)$. 

It is then convenient, as we want to work exclusively on the 
boundary, to adopt a definition of scalar product
\ba
\ll \phi | \psi \rr ~ \equiv ~{1 \over {2 k^{2}}}
~\oint~ds~{\bf \hat n} \cdot {\bf r}~
\phi^{*}(s) \psi(s).
\ea

Any of these functions can be extended to the domain by means 
of Green's theorem;
using it as an ansatz for the normal derivative, 
${ {\partial \Psi} \over {\partial {\bf n}} }(s)$, and 
setting $\Psi(s)$ to zero, 
\ba
\label{green2}
\Psi_{k} ({\bf r})= 
\oint ds ~ \left[
\Psi ({\bf r'}(s) ) ~
{ {\partial G_{0} } \over {\partial {\bf n}} } 
(k;{\bf r},{\bf r'}(s)) ~-
~ G_{0}(k;{\bf r},{\bf r'}(s)) ~
{ {\partial \Psi } \over {\partial {\bf n}} } ({\bf r'}(s) )
\right].
\ea
Of course, (\ref{green2}) is not an 
eigensolution, because the limiting value of $\Psi({\bf r})$, as 
${\bf r}$ goes to the boundary, is not zero; that is to say, the 
function is discontinuous at the boundary. This function 
depends on the continuous parameter $k$, which controls the 
semiclassical limit.

With this definition, the normal derivatives $\phi_{\nu}(s)$ of 
eigenfunctions normalized {\it in the domain} are orthonormal (to order 
$1/k$) in a small range $\Delta k=(2 \times~\mbox{perimeter/area})$.

For the phase space representation we construct 
coherent states with the correct space periodicity \cite{voros}, 
defined as 
\ba
\langle s|p~q \rangle=
\left({k \over {\sigma \pi}}\right)^{1/4}
\sum_{a=-\infty}^{\infty}
 \exp\left( {i~k~p~(s-a)} \right)
\exp \left( {-{ k \over {2 \sigma}} (s-q-a)^{2}} \right).  
\ea
This is a boundary wave packet, periodic in $s$, which is localized at 
the point $(p,q)$ in the Birkhoff Poincar\'e section phase space.

A single wave packet represents a bounce off a specified point on 
the boundary with a given tangential momentum. Thus, to extract 
the phase space contents of a given eigenfunction, we can 
construct the overlap 
 \ba
{\cal A}_{\nu}(p,q)~=~ 
{1 \over {| \ll p~q|p~q \rr |^{1/2}}}
\oint  ds ~ 
\langle p~q|s \rangle ~ \phi_{\nu}(s) ~ 
{ {\bf \hat n \cdot \vec r} \over {2 k_{\nu}^{2}} }.
\ea

Thus, a first visual display of the eventual
localization and scarring of the eigenstates comes through the Husimi
function
\ba
\label{hhh}
{\cal H}_{\nu}(p,q)~=~ \left| {\cal A}_{\nu}(p,q) \right|^{2}.
\ea
We show an example of this for the stadium eigenfunction with
$k=100.954920427642$ in Figure \ref{husimi1}.

Clearly, a first quantitative measure of the scarring of a 
periodic orbit would be
\ba
\label{sprime}
S'(\nu,\gamma)= {1 \over N} \sum_{i=1}^{N} {\cal H}_{\nu}
\left(p_{\gamma_i},q_{\gamma_i} \right)
\ea
which averages the probability over the $N$ points $\gamma_{i}$ 
of the periodic trajectory $\gamma$. This measure was used by 
Muller and Wintgen in the context of the diamagnetic Kepler problem 
\cite{muller}.

This average over the probabilities of the periodic points does not 
take into account the phase relations, due to semiclassical 
propagation, beetween them. 
Therefore, it seems a more convenient strategy to average the 
{\it amplitudes} with the proper phase differences.
A better measure, then, is provided by the construction of the 
scar function 
\ba
\label{pecstate}
\langle s|\varphi (k,\gamma) \rangle ~ = ~ 
{1 \over {|\ll\varphi (k,\gamma ) |\varphi (k,\gamma ) \rr |^{1/2} } }
 \sum_{j=1}^{N} \exp \left( i f_{j} \right)
\langle s |p_{j}q_{j} \rangle
\ea
where $q_{i}$ and $p_{i}$ are the Birkhoff coordinates of the 
periodic points.

The phases $f_{j}$ are defined by
\ba
f_{j}=k l_{j} - j \pi - {\pi \over 2} \nu_{j}
\ea
where $l_{j}$ is the distance in configuration space beetween the 
initial point of the periodic orbit and the $j-th$ point. The second
term takes into account the boundary conditions (Dirichlet) and the 
third, the conjugate points along the trajectory.
The inclusion of these phases is very important in the 
determination of the existence of scars.

The total accumulated phase, $f_{N}$, will not be, in general, a 
multiple of $2 \pi$. In order to have an invariant function, 
depending only on the orbit and not on the starting point, we add 
an additional phase to each point, so as to make $f_{N}=2 \pi n$,  
with $n$ an integer:
\ba
f_{j} \rightarrow f_{j} + {j \over N} \alpha
\ea
with $\alpha$ the minimum beetween $(f_{N})_{\mbox{mod}~2 \pi}$ and 
$2 \pi -(f_{N})_{\mbox{mod}~2 \pi}$.

This state is a coherent sum over a periodic orbit and, thus, is a 
good candidate for an invariant probe depending only on the  
orbit. 

\subsection{The Scar Intensity $S_{\gamma}(k_{\nu})$ and the 
Scar Length Spectrum ${\tilde S}_{\gamma}(l)$ }

We define the scar intensity
\ba
\label{ssss}
S(\nu,\gamma)= |\ll \varphi(k_{\nu},\gamma)|\phi_{\nu} \rr|^{2}.
\ea
Notice that $k$ is set to $k_{\nu}$ in $\langle 
\varphi(k,\gamma)|$.
This measure of the scar intensity differs from (\ref{sprime}) 
mainly by interference terms.  

Each wave packet in (\ref{pecstate}) represents a localized plane 
wave hitting the boundary at a specified point in a specified 
direction. Thus (\ref{pecstate}), when 
seen in this light,  can be assimilated to a superposition of 
plane waves which privileges the wave directions associated to 
the periodic orbit. 
For example, in Figure \ref{figu} we show a periodic orbit of 
the stadium billiard 
and the associated scar wave function in different representations, 
with $k=100.954920427642$ (the same as in Figure \ref{husimi1}).

The scar wave function in the domain is a solution of the 
Helmholtz equation with a given value of $k$. 
So, if we expand it in terms of the exact eigenfunctions, we expect 
that the more significant contributions come from the eigenfunctions 
with closer $k_{\nu}$ to $k$ (more precisely, 
$|k-k_{\nu}|\leq (2 \times~\mbox{perimeter/area})$.
Then, using the quasi-orthogonality relation (\ref{qortho})
the norm of the scar wave function in the domain is $1$ to order 
$k^{-1}$.
Our aim is to describe this subspace in terms of states constructed 
on periodic orbits, as in (\ref{ssss}).
This is not dissimilar to the task of describing them in terms 
of plane or cylindrical waves. However, the peculiar linear 
combinations taken in (\ref{pecstate}), 
being semiclassically invariant under the bounce map, 
should provide the most important correlations.

The density of states of the billiard has a semiclassical 
representation \cite{smilden} as 
\ba
\label{semiden}
d_{scl}(k)~ \approx ~\langle d(k) \rangle ~+~
{1 \over {\pi}} ~ \sum_{p}~\sum_{r=1}^{\infty}~
{{l_{p}} \over {|\mbox{det}({\bf I}-{\bf T}_{p}^{r})|^{(1/2)} }}~
\cos \left( r ( k l_{p} - \nu_{p} \pi/2 ) \right)
\ea
where the first sum is done over primitive periodic orbits, the 
second sum takes into account their repetitions; $l_{p}$ is the 
length of the orbit, $\nu_{p}$ is the Maslov index, ${\bf T}_{p}$ 
is the monodromy matrix. The smooth part of the density of states 
is given by $\langle d(k) \rangle$.

The Fourier transform of (\ref{semiden}) provides a distribution 
linked more directely 
to the classical motion, i. e., the {\it length spectrum}, which 
shows well defined peaks at the lengths (actions) of periodic 
orbits.

In order to focus more specifically on the scarring features of a 
single orbit along the spectrum, we set $\gamma$ to a given 
periodic orbit in (\ref{ssss})
\ba
\label{sint}
S_{\gamma}(k_{\nu})= |\ll \varphi_{\gamma}(k_{\nu})|\phi_{\nu} \rr|^{2}.
\ea
Its Fourier transform is 
\ba
\label{lspec}
{\tilde S}_{\gamma}(l)= \sum_{k_{\nu}} 
S_{\gamma}(k_{\nu})~\exp \left( i k_{\nu} l \right),
\ea
which we will call the scar length spectrum.

If periodic orbits obeyed Bohr-Sommerfeld like quantization 
conditions, we would expect a periodic behaviour of 
$S_{\gamma}(k_{\nu})$ with a period $\Delta k=\pi /L_{\gamma}$.
This implies periodic sequences of scarred states along the 
spectrum . These states  
have been also observed in other billiards \cite{agam} \cite{smilansky}. 
Here we test for these periodicities directly in the scar length 
spectrum.
The periodicities would be exact but for the fact that the periodic 
orbit basis is not orthonormal, and the quantization rule of a single 
orbit does not lead necessarily to a quantized state.

To eliminate spurious behaviour of the Fourier transform due to end 
effects of the $k$-interval, we multiply the scar intensity in 
(\ref{lspec}) by a function vanishing at the ends; 
typically a quadratic function \cite{prosen}.
The scar length spectrum still shows large fluctuations with very 
clear local average peaks at certain lengths. This local structure is evidenced by averaging the resulting Fourier transform.

\section{The Bunimovich Stadium}
\label{bunis}

The boundary of the stadium billiard is defined by two 
semicircunferences connected 
by two straight segments. Of all the possible stadia, 
we consider only the one with relation $2:1$ beetween its total 
length and height, and we will scale the lengths in such a way  
that the perimeter is $4+2 \pi$ and the area is $4+ \pi$.

At the classical level, we will describe the periodic orbits 
following Biham and Kvale \cite{biham}. Their symbolic dynamics is 
a six-symbol one where each symbol corresponds to a bounce off the 
boundary:
\begin{itemize}
\item 0: A bounce off the lower straight segment.
\item 1: A clockwise bounce off the left semicircle or a single 
anticlockwise bounce off the left semicircle.
\item 2: A bounce off the upper straight segment.
\item 3: A anticlockwise bounce off the right semicircle or a single 
clockwise bounce off the left semicircle.
\item 4: A not single anticlockwise bounce off the left semicircle.
\item 5: A not single clockwise bounce off the right semicircle.
\end{itemize}
A bounce is a single one if it is not preceded or followed by a 
bounce off the same section of the boundary.

This dynamics has to be pruned. This pruning is geometrical and corresponds 
to the symbolic dynamics of the stadium of infinte length. As the 
length is made finite, more pruning rules appear (of a dynamical 
character) as described in \cite{cvitanovic}.

Using this symbolic description 
we have computed all the periodic orbits up to 10 bounces 
and a few selected ones of much higher periods, and we have ordered them 
(somewhat arbitrarily) by their number of bounces and symbolic codes.
In Figure \ref{somepos} we show the correspondence of these calculated 
orbits as projections in configuration space and in the Birkhoff 
Poincar\'e (desymmetrized) section.  

At the quantum level, we compute the energy levels and eigenfunctions 
by the scaling method \cite{vergini}, which gives directly all 
eigenvalues and eigenfunctions very precisely and efficientely. 
We have computed 1654 consecutive levels and their eigenfunctions 
ranging from $k \approx 62.8$ and $k \approx 125.2$; 
and other selected ones.

\subsection{Periodic Orbit Decomposition of Eigenfunctions}

In Figure \ref{wwphases} we demostrate the advantage of using 
the present scar function as opposed to the simple unphased 
average, $S'$, of (\ref{sprime}). We take one of the scarred 
wavefunctions of Heller \cite{heller} (Figure \ref{hellerscars}) 
and plot the quantities $S$  and $S'$ as a function of the periodic 
orbit label $\gamma$ (Figure \ref{wwphases}). The periodic orbits 
are ordered in increasing periods and, within each period, 
by symbolic codes. Recurrences in $S$ sometimes occur due to the 
existence of orbits of period $n \times p$ that almost retrace 
$n$ times the periodic orbit of period $p$. However, many 
recurrences are also due to short homoclinic and heteroclinic 
excursions.

The peaks are more clearly defined in the $S$ plot, due to the 
enhancement brought about by the semiclassical dynamics $f_{j}$ 
(note that 
the maximum scar value is around 0.114 for $S$ and 0.06 for $S'$). 
Due to the binning of the interval, the strongest peak of $S'$ 
overlaps with one of the secondary peaks of $S$.
In the leftmost inset in Figure \ref{wwphases} we show the 
periodic orbit that scars this eigenfunction the most. Moreover, 
the two most prominent secondary peaks come from two different 
homoclinic excursions of this orbit (see rightmost inset in 
Figure \ref{wwphases}).

Both measures indicate the presence of scars, i.e., amplitudes larger 
than the average fluctuation. However, as $S$ captures the phase 
relations of periodic orbits, the basis $| \varphi(k,\gamma) \rangle$ 
is `closer' to reflect the invariant properties characteristic of the 
stadium. Thus, we expect the amplitudes $S$ to have much larger 
fluctuations (and, therefore, clearer scars) than $S'$.
This fact shows clearly in Figure \ref{binscsf}, where we show the 
distribution of scar intensities, $N(S)$. The small amplitudes are 
distributed approximately as an exponential. The distribution 
corresponding to $S$ is much broader than that of $S'$. 
For this strongly scarred state, there is a large region beetween 
$0.07$ and $0.114$ where no scar intensities appear. So, 
$\mbox{log}_{10}(N(S))$ goes to $-\infty$ in this region.
At $0.114$ a single periodic orbit gives a large scar, yielding 
$\mbox{log}_{10}(N(S))=0$. The occurrence 
of this peak is, however, a rare event. The secondary peaks, due 
to homoclinic excursions of the periodic orbit which gives the 
strongest peak, are to be found at, approximately, $S=0.071$ and 
$S=0.066$.

We have tested all 1654 
eigenfunctions in the range beetween $k \approx 62.8$ and 
$k \approx 125.2$
against scarring by the first 617 periodic orbits (i.e., up to 
9 bounces), giving a total of approximately 1 million scar intensities. 
The distribution of these intensities is shown in Figure \ref{allbins}. 
We can observe three different sections: strong scars, weak scars 
and the region closer to $S=0$, where most of the scar intensities are 
(there are around a million scar intensities from $S=0$ to $S=0.1$, 
whereas only 760 are to be found with $S>0.1$.) This distribution is 
very different from the Porter-Thomas result. The reason is that the 
basis of $\langle s|\varphi (k,\gamma) \rangle$ is not orthonormal 
and, moreover, is chosen so as to be closely related to the 
dynamics.
Should we examine the scars in any other basis, unrelated to the 
dynamics of the stadium, for example a plane wave basis, we would only 
expect a statistical distribution of the intensities, in accordance 
to random matrix theories.

The presence of the peaks in the strong scar region in Figure 
\ref{allbins} quantifies the scar phenomenon and shows, in 
accordance to Shnirelman's theorem \cite{shnirelman}, 
that scarring is exceptional. 
However, it is the only remaining signature of the specific classical 
behaviour of the system, as embodied in its periodic orbits. 

\subsection{Families of Periodic Orbits}
 
\subsubsection{Whispering Gallery Family}

This family is composed by trajectories 
of code $5^{n}0^{a}1^{m}2^{b}$; where $a$ and $b$ are 0 or 1, and 
$m$ and $n$ are positive integers 
(as there is time reversal symmetry, the same orbit can be described 
by the code $3^{n}2^{b}4^{m}0^{a}$). The whispering gallery limit 
is approached as $m,n \rightarrow \infty$ simultaneously. For the 
periodic orbit to exist, as this limit is approached, the 
difference beetween $m$ and $n$ should remain finite; the largeness 
of this difference being determined by the length of the stadium.
This is a clear example of dynamical pruning (as opposed to 
geometrical pruning, i.e. independent of the length of the stadium)
 
We have found that the whispering gallery trajectories
that show a more pronounced periodicity in the scarring are those 
defined by $m>2$ and $n>2$, independent of their symmetry or 
value of $a$ and $b$. Of course, where high symmetry is present, 
the periodicity is stronger. 

We show in 
Figure \ref{somewg} the function $S_{\gamma}(k)$ and the scar length 
spectrum for some whispering gallery periodic orbits.
We see how the scar length spectrum shows clearly the periodicity 
of $S_{\gamma}(k)$, 
defined by the fundamental length of the orbit and its repetitions. 
(The lengths shown are multiples of $L_{\gamma}/2$ because of the 
symmetry of the orbit.)

The width of the group of states that participate in the scarring 
is constant in the region considered. This means that more and more 
states are involved in one `Bohr-Sommerfeld' interval. However, 
only a few, typicaly one or two, show visible scars. In Figure 
\ref{somewg} (first panel), 
where the width of the groups is approximately $\Delta k \approx 1.27$, 
the first group involves around 35 states at $k \approx 62.88$ 
and the last involves 45 at $k \approx 125.60$ in accord to the 
change in the density of states.

We can exemplify how well the scar intensity picks up the scarred 
eigenstates. We look for scars in the region delimited by 
$k \approx 77.14$ and $k \approx 77.34$
for the periodic orbit with code $55551111$ (see Figure \ref{sc235}).
We find two contiguous 
states that are scarred by this periodic orbit, namely 
$k=77.22171991174$ and $k=77.24033835210$. The probability densities of 
these eigenfunctions confirm this fact: see Figure \ref{eigens235}. 
Notice the similarity beetween both 
eigenfunctions and beetween them and the mentioned whispering gallery 
periodic orbit. 

\subsubsection{Bouncing Ball Family}

This family is composed of three subfamilies, following \cite{tanner}, 
whose symbolic codes are 
$33(02)^{n}11(02)^{n}$ (family $A$), 
$23(20)^{n}21(20)^{n}$ (family $B$), 
$3(20)^{n}1(02)^{n}$ (family $C$).
The bouncing ball limit is $n \rightarrow \infty$, where the resulting 
periodic orbit has increasingly smaller $x$ component of the 
wavenumber $k$. We will consider the first five members of each family, 
with periods ranging from 6 to 24 (see Figure \ref{bballpo}).

The bouncing ball eigenstates are approximately described by those of 
a rectangle with the same size as the one inscribed in the stadium. 
So, the quantized wavenumbers are given by 
$k \approx  \pi \sqrt{n_{x}^{2}+n_{y}^{2}}$, where $n_{x(y)}$ is the 
number of nodes along the $x(y)$ axis.

The preceding considerations tell us not to expect a scar length 
spectrum that is peaked in the length 
of the given periodic orbit and its multiples. This is so 
because there is no such periodicity in the scar intensities for 
the bouncing ball families. This is what we observe for those 
orbits tending to the 
bouncing ball limit, see an example in Figure \ref{bblimit}. 
Notice, though, how for big $L$ the peaks appear for even $L$. 
This fact is related to the multiple bounces beetween the two 
straight segments of the billiard, approximately of length $2$.
Some of the first few orbits of 
each family show a single peak in the length of the orbit, with 
no peaks (or small ones) in the multiples, see an example in 
Figure \ref{nobblimit}.

\subsubsection{Other Periodic Orbits}

As `scars are scarce' \cite{smilansky}, the scar length spectrum 
for most periodic orbits shows no peaks in the associated length. 
Other orbits show rather more complex patterns with many lengths 
that are not easily assigned to other periodic orbits (whether 
in the homoclinic family of the first or not).
However, there is a small set of orbits for which the scar intensities 
have the expected periodicity (some examples in Figure \ref{pogood}).

\section{Conclusions}

We have constructed a quantitative measure for the presence of scars 
that, by taking into account semiclassical phase correlations, 
provides a sharper indication of their presence. The measure is 
constructed as an ansatz for the normal derivative of a state 
representing a pure stationary scar and therefore testing the most 
the irreducible contents of the eigenfunctions as contained in 
the normal derivative. Thus, it is 
very suitable for numerical calculations, 
as it involves only boundary integrations, avoding completely- 
except for graphical display- any integration over the domain 
of the billiard. 

We have provided examples of the decomposition of a single 
eigenstate into scar functions and of the systematic way in which 
some orbits appear in the $k$ spectrum. The scar intensities show 
how an eigenstate is distributed on a basis of quantum states 
constructed from periodic orbits. Such a basis is clearly not 
orthogonal (and, probably, overcomplete). The investigation of 
the properties of this basis remains to be done.

By testing many orbits and many eigenstates we have found that 
scars are quite rare, 
in accordance to expectations from Shnirelman's theorem 
\cite{shnirelman}.
Even less frequent is to find sequences of states scarred periodically 
(in $k$) by a given orbit. 
Most eigenfunctions decompose in periodic orbits in 
such a way that no one prevails over the others. This, in turn, 
implies no clear scars in most instances.

The families that show stronger scars are the whispering gallery 
and the bouncing ball ones, both being rather exceptional families.

\section*{Acknowledgements}

This work is partially supported by CONICET PIA 6950 and by the 
EC Programme/ARG/B7-3011/94/27. We are grateful to the referee for 
valuable suggestions in improving the presentation.



\newpage

\section*{Figure Captions}

\begin{figure}[h] 
\caption[]{\label{husimi1} Eigenfunction (left panel) and Husimi
 representation (right panel) for $k=100.954920427642$. This function 
is scarred by the periodic orbit with code $23202120$. The crosses 
are the periodic points of this orbit.}
\end{figure}

\begin{figure}[h] 
\caption[]{\label{figu} 
Different representations of the scar function of a periodic orbit.

Upper left panel: Periodic orbit in the fundamental domain
(symbolic code 23202120, following \cite{biham}) 

Upper right panel: Real part of the scar wave function in the boundary.

Lower left panel: Probability density in the domain 
(via Green function).

Lower right panel: Husimi representation.}
\end{figure}

\begin{figure}[h] 
\caption[]{\label{somepos}Periodic orbits of 4 bounces in the 
configuration space (upper panel) and in phase space (lower panel)
In the latter plot, the points 
are depicted by dots or crosses, depending on the sign of $p$. The 
dotted vertical line is placed at the value of $q$ where the 
discotinuity in curvature occurs.}
\end{figure}

\begin{figure}[h] 
\caption[]{\label{hellerscars} Odd-odd stadium eigenfunction with 
$k=130.4886755073$.}
\end{figure}

\begin{figure}[h] 
\caption[]{\label{wwphases} $S$ and $S'$ functions for k=130.4886755073 
(upper and lower panel, respectively.)
In the leftmost inset we see the periodic orbit with maximum scar 
intensity; in the rightmost one, the orbits that give the two secondary 
peaks. The orbits are ordered by increasing number of bounces, up to 
9.}
\end{figure}

\begin{figure}[h] 
\caption[]{\label{binscsf} Distribution of scar intensities, $N(S)$, for 
$S$ (full line) and $S'$ (dashed line) for k=130.4886755073.} 
The base of the logarithm is $10$.
\end{figure}

\begin{figure}[h] 
\caption[]{\label{allbins} Distribution of scar intensities, $N(S)$, for 
the 1654 consecutive eigenfunctions and first 617 periodic orbits.}
The base of the logarithm is $10$.
\end{figure}

\begin{figure}[h]
\caption[]{\label{somewg} Scar intensity and scar length spectrum for 
some whispering gallery periodic orbits. 

Upper left panel: Scar intensity for periodic orbit 3332440.

Lower left panel: Scar length spectrum for periodic orbit 3332440.

Upper right panel: Scar intensity for periodic orbit 55511111.

Lower right panel: Scar length spectrum for periodic orbit 55511111.}
\end{figure}

\begin{figure}[h]
\caption[]{\label{sc235} Scar intensity for a range of $k$, 
for the periodic orbit depicted in the inset.}
\end{figure}

\begin{figure}[h]
\caption[]{\label{eigens235} Group of eigenfunctions in the 
same range of $k$ of Figure \ref{sc235}. }
\end{figure}

\begin{figure}[h]
\caption[]{\label{bballpo} Bouncing ball periodic orbits, first 
five members of familes A, B and C (first, second and third row, 
repectively)}
\end{figure}

\begin{figure}[h]
\caption[]{\label{bblimit} Scar intensity and scar length spectrum 
for a periodic orbit in the bouncing ball limit.}
\end{figure}

\begin{figure}[h]
\caption[]{\label{nobblimit} Scar intensity and scar length spectrum 
for a periodic orbit not in the bouncing ball limit.}
\end{figure}

\begin{figure}[h]
\caption[]{\label{pogood} Scar intensity and scar 
length spectrum for some low period periodic orbits.

Left panels: Scar intensity and scar length spectrum for periodic 
orbit 3210.

Right panels: Scar intensity and scar length spectrum for periodic 
orbit 2321.}
\end{figure}


\newpage

\begin{thebibliography}{99}

\bibitem{mcdonald} McDonald S. W., {\it PhD Thesis}, Lawrence Berkeley 
Laboratory LBL 14837 (1983)

\bibitem{heller} Heller E. J., in {\it Chaos and Quantum Physics} 
(Proceedings from Les Houches 1989)

\bibitem{bogoscars} Bogomolny E. B., Physica D {\bf 31} (1988), 
169-189.

\bibitem{berryscars} Berry M., Proc. Roy. Soc. A {\bf 423} (1989), 219.

\bibitem{agam} Agam O. and Fishman S.; Phys. Rev. Lett. Vol 73, 
6 (1994), 806-809.

\bibitem{smilansky} Klakow D. and Smilansky U.; J. Phys. A {\bf 29}, 3213
(1996)

\bibitem{voros} Tualle J. M. and Voros A., Chaos, Solitons and 
Fractals, Vol. 5 Nr. 7, 1085 (1995)

\bibitem{muller} Muller K. and Wintgen D., J. Phys. B,
Vol. 27 Nr. 13, 2693 (1994)

\bibitem{berry} Berry M. and Wilkinson M., Proc. Roy. Soc. A {\bf 392}, 
15-42 (1984)

\bibitem{arnold} Arnold V. I. and Avez A., {\it Ergodic Problems of 
Classical Mechanics}, Addison Wesley (1989).

\bibitem{vergini} Vergini E. and Saraceno M., Phys. Rev. E {\bf 52}, 2204 (1995)

\bibitem{smilden} Smilansky U., in {\it Mesoscopic Quantum Physics} 
(Proceedings from Les Houches 1994) 

\bibitem{prosen} Prosen T., preprint chao-dyn/9611015

\bibitem{biham} Biham O. and Kvale M., Phys. Rev. A {\bf 46}, 6334 (1992)

\bibitem{cvitanovic} Hansen K. T. and Cvitanovic P., 
preprint chao-dyn/9502005

\bibitem{shnirelman} Shnirelman A., Usp. Mat. Nauk. {\bf 29} (1974), 181

\bibitem{tanner} Tanner G., preprint chao-dyn/9610013


\end{thebibliography}
\end{document}